\documentclass{llncs}

\usepackage{graphicx}
\usepackage{listings}
\usepackage[utf8]{inputenc}
\usepackage{comment}
\usepackage{showexpl}

\usepackage{epsfig}
\usepackage{calc}
\usepackage{amssymb}
\usepackage{amstext}
\usepackage{amsmath}
\usepackage{enumitem}
\usepackage{pslatex}
\usepackage{multirow}

\usepackage{color}
\usepackage[hyphens]{url}
\newcommand{\kari}      [1] {\noindent \textcolor{blue}{[#1 -Kari]}}

\definecolor{mygreen}{rgb}{0,0.6,0}
\definecolor{mygray}{rgb}{0.5,0.5,0.5}
\definecolor{mymauve}{rgb}{0.58,0,0.82}

\lstdefinestyle{frame-number}{
  style=basic-style,
  frame=single,
  numbers=left,
  numbersep=1ex,
  xleftmargin=5.0ex,
  framexleftmargin=3ex
  }

\begin{document}
% Added by farshad to break a url into sevral line
%as suggested here: https://stackoverflow.com/questions/2640111/url-latex-linebreak
\sloppy
%\title{A Decomposition and Metric-Based Evaluation Framework for Microservices\\OR\\How to Decompose a Monolithic System into Microservices and how to Evaluate the Decomposition\\OR\\Change from From Monolithic Systems to Microservices: A Decomposition Framework based on Process Mining}
\title{A Decomposition and Metric-Based Evaluation Framework for Microservices}
%
%\titlerunning{Hamiltonian Mechanics}  % abbreviated title (for running head)
\author{Davide Taibi, Kari Syst{\"a}}
%\author{\authorname{Davide Taibi\orcidAuthor{0000-0002-3210-3990}, Kari Syst{\"a}\orcidAuthor{0000-0001-7371-0773}}}
%
%\authorrunning{Ivar Ekeland et al.} % abbreviated author list (for running head)
\tocauthor{Davide Taibi, Kari Syst{\"a}}
\institute{TASE - Tampere Software Engineering Research Group \\ Tampere University. Tampere Finland\\
\email{[davide.taibi;kari.systa]@tuni.fi}}

\maketitle              % typeset the title of the contribution


\begin{abstract}
Migrating from monolithic systems into microservice is a very complex task. Companies are commonly decomposing the monolithic system manually, analyzing dependencies of the monolith and then assessing different decomposition options. The goal of our work is two-folded: 1) we provide a microservice measurement framework to objectively evaluate and compare the quality  of microservices-based systems; 2) we propose a decomposition system based on business process mining. 
The microservice measurement framework can be applied independently from the decomposition process adopted, but is also useful to continuously evaluate the architectural evolution of a system. 
Results show that the decomposition framework helps companies to easily identify the different decomposition options. The measurement framework can help to decrease the subjectivity of the decision between different decomposition options and to evaluate architectural erosion in existing systems. 
\end{abstract}

\section{Introduction}
\label{sec:introduction}

Software evolves through its life-time, and often a large part the effort and costs is spent on software maintenance \cite{Lehman1980}. Furthermore, the incremental development practices in modern software development makes the nature of all the development processes to resemble a maintenance  one \cite{TommiKari2014}.
A major obstacle to efficient maintenance is the tight coupling between the internal components of the software. In monolithic systems, most changes require modifications to several parts of the systems, and often size and complexity of the modification is hard to estimate in advance.

One approach to tackle the maintenance problem is to decompose the system into small and independent modules \cite{Parnas1972} \cite{SOLDANI2018}. Often, at the same time, companies want to utilize benefits of service-oriented architectures and even \textit{microservices}, such as independent development, scaling and deployment ~\cite{TaibiIEEECloud}.
%Legacy and monolithic system have become hard to maintain because of tight coupling between their internal components. Modifying a feature in one class often involves changes in several other classes, thereby increasing the needed development time and effort. The decomposition into small and independent modules is a strategy that companies may adopt to improve maintainability~\cite{Parnas1972} \cite{SOLDANI2018}. Often, at the same time, companies want to utilize benefits of microservices, such as independent development, scaling and deployment ~\cite{TaibiIEEECloud}. 

Microservices are an adaptation service-oriented architecture but focuses on of relatively small and independently deployed services, with a single and clearly defined purpose~\cite{fowler2014microservices}. The independent development and deployment bring several advantages. Different microservices can be developed in different programming languages, they can scale independently from other services, and each microservice can be deployed on the most suitable the hardware. Moreover, small services are easier to maintain and the split to independent responsibilities increases fault-tolerant since a failure of one service will not break the whole system. From the architectural perspective a well-designed microservice encapsulates its data and design choices. Thus, the internal logic of a microservice can be changed without affecting the external interface. This reduces the need for interaction between the teams~\cite{TaibiCommunication}~\cite{TaibiRequirements}.
%Microservices are relatively small and autonomous services deployed independently, with a single and clearly defined purpose~\cite{fowler2014microservices}. Because of their independent deployment, they have a lot of advantages. They can be developed in different programming languages, they can scale independently from other services, and they can be deployed on the hardware that best suits their needs. Moreover, because of their size, they are easier to maintain and more fault-tolerant since a failure of one service will not break the whole system, which could happen in a monolithic system. Since every microservice has its own context and set of code, each microservice can change its entire logic from the inside, but from the outside, it still does the same thing, reducing the need of interaction between teams~\cite{TaibiCommunication}~\cite{TaibiRequirements}.

However, decomposing a monolithic system into a set of independent microservices is a very difficult and complex tasks ~\cite{TaibiIEEECloud}\cite{TaibiXP17}.
Decomposition of a system into separately maintained services is difficult as such, but microservice architecture adds further challenges related to performance. Calls inside a microservice are significantly lighter than calls between microservices.
Still, the quality of the decomposition -- the optimal slicing of the monolith to services -- is critical for gaining the assumed benefits of using microservices. The software architects usually perform the decomposition manually but the practitioners have claimed that a tool to support identification different possible slicing solutions~\cite{SOLDANI2018}\cite{TaibiIEEECloud}~\cite{TaibiCLOSER} would greatly help the task.  Typically, the only helping tools for the software architects have been based on the static analysis of dependencies such as Structure 101\footnote{Structure101 Software Architecture Environment - http://www.structure101.com}. The actual discovery of slicing options has still been done by experienced software architects.
In microservices, the dynamic behavior of the system is important too since it affects the performance and maintainability. Since static dependency analysis tools are not able to capture the dynamic behavior we decided to explore slicing based on runtime behavior instead of only considering static dependencies.
%However, decomposing a monolithic system into independent microservices is one of the most critical and complex tasks ~\cite{TaibiIEEECloud}\cite{TaibiXP17} and several practitioners claim the need for a tool to support them during the slicing phase in order to identify different possible slicing solutions~\cite{TaibiIEEECloud}~\cite{TaibiCLOSER}. The decomposition is usually performed manually by software architects ~\cite{TaibiIEEECloud}\cite{SOLDANI2018}. Up to now, the only help that software architects can have is based on the static analysis of dependencies with tools such as Structure 101\footnote{Structure101 Software Architecture Environment - http://www.structure101.com} while the slicing of the system commonly is delegated to the experience of the software architect itself. Moreover, static dependency analysis tools are not able to capture the dynamic behavior of the system and run-time dependencies like frequent method calls could have an influence to both maintainability and performance. Thus, we decided to approach the slicing based on runtime behavior instead of only considering static dependencies.

In our previous work work, we proposed a  microservice decomposition framework~\cite{DavideKari2019} based on process-mining to ease the identification of splitting candidates for decomposing a monolithic system into separate microservices. The framework is based on logs produced by process mining of the original monolithic system. 
% We assume that the software architects make the actual decision on decomposition, but the proposed tool can create proposal and provide the architects with data about the consequences of the decomposition. This reduces the subjectivity and the risks of related to the slicing decomposition.
%In order to ease the identification of microservices in monolithic applications, we adopted a data-driven approach for identifying microservices candidates based on process mining performed on log files collected at runtime. Our approach combines process mining techniques and dependency analysis to recommend alternative slicing solutions. Our decomposition approach can be used by software architects to support their decisions and to help them easily identify the different business processes in their applications and their dependencies, reducing the subjectivity and the risks of related to the slicing process.
The decomposition framework has been also validated in our previous study~\cite{DavideKari2019} in collaboration with an SME. 
% We supported company in the migration to microservices and in comparing the proposed decomposition solution with proposal by the software architect. 
% We also collected feedback from the stakeholders to learn about practical applicability of the approach.
%We validated this work with an industrial case study performed in collaboration with an SME that we supported in the migration phase, comparing the decomposition solution proposed by the software architect with the one obtained from the application of our process-mining based approach.
The results of~\cite{DavideKari2019}  shows that dynamic call history can be effectively used in the decomposition of microservices.  This approach can also identify  architectural issues in monolithic systems.
The approach can be used by companies to proposed different slicing options to the software architects and to provide additional analysis of the software asset.
This would reduce the risk of wrong slicing solutions.
%The results show that process mining can be effectively used to support the decomposition of microservices and that it also supports the identification of existing architectural issues in monolithic systems. The result can be used by companies to reduce the risk of a wrong slicing solution, suggesting different slicing options to the software architects and providing additional analysis of the software asset.

In this paper, we extend the previous decomposition framework~\cite{DavideKari2019} proposing a new measurement framework to objectively compare two decomposition options. 
The measurement framework  can be used  independently from the decomposition strategy adopted. 

% In this version we extend the discussion and analysis of the role of dynamic behavior of the slicing, with a more detailed   description of the approach and practical implementation as future work. 

The remainder of this paper is structured as follows. 
Section \ref{sec:background} presents the background on processes for migrating and splitting monolithic systems into microservices. Section \ref{sec:measurementFramework} presents the measurement framework. Section \ref{sec:Framework} describes our proposed decomposition approach.
%Section 5 reports on the industrial case study.
Section \ref{sec:discussion} discusses the results. Section \ref{sec:related} presents related works while, finally Section \ref{sec:conclusions} draws conclusions.

\section{Background and Assumptions}
\label{sec:background}
Decomposing a system into independent subsystems is a task that has been performed for years in software engineering. Parnas~\cite{Parnas1972} proposed the first approach for modularizing systems in 1972. After Parnas’s proposal, several works proposed different approaches~\cite{LenarduzziICSE2017}. Recently, the decomposition of systems took on another dimension thanks to cloud-native systems and especially microservices. In microservices, every module is developed as an independent and self-contained service. 

\subsection{Microservices}

Microservices are small and autonomous services deployed independently, with a single and clearly defined purpose~\cite{fowler2014microservices},~\cite{Newman2015}. 
 In microservices  each service can be developed using different languages and frameworks. Each service is deployed to their dedicated environment whatever efficient for them. 
 
 The communication between the services can be based on either REST or message queue. So, whenever there is a change in business logic in any of the services, the others are not affected as long as the communication endpoint is not changed. As a result if any of the components of the system fails, the failure will not affect the other components or services, which is a big drawback of monolithic system~\cite{fowler2014microservices}. 
 
\begin{figure*}[!ht]
\vspace{6mm}
\centering 
\includegraphics[trim= 0 50 0 0,  clip, width=1\linewidth]{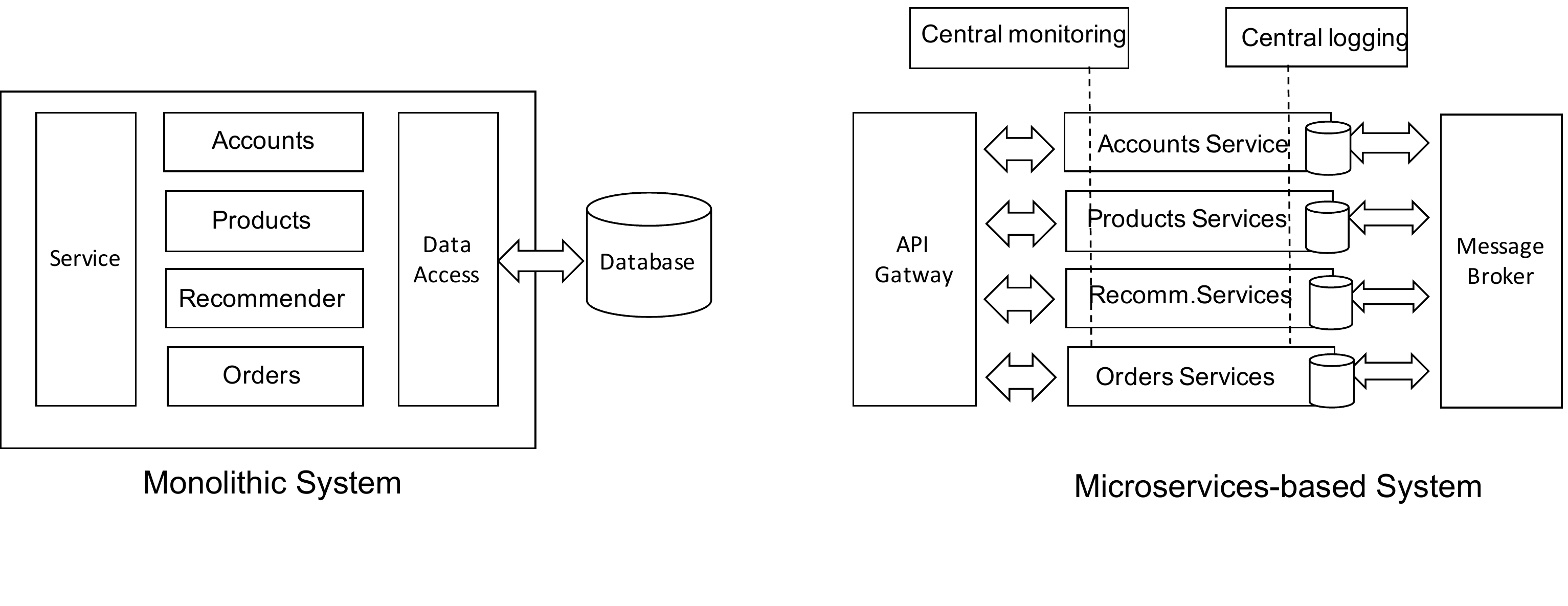}
\caption{Architectures of Microservices and Monolithic systems}
\label{fig:Microservice}
\vspace{6mm}
\end{figure*}

 As  we can see in Figure \ref{fig:Microservice},  components in monolithic systems are tightly coupled with each other so that failure of one component will affect the whole system. Also if there is any architectural changes in a monolithic system it will also affect other components. Due to these advantages, microservice architecture is way more effective and efficient than monolithic systems. 

\subsection{The Microservice Decomposition Process}

Taibi et al.~\cite{TaibiIEEECloud} conducted a survey among 21 practitioners who adopted microservices at least two years ago. The aim of the survey was to collect their motivation for, as well as the evolution, benefits, and issues of the adoption of microservices. Based on the results, they proposed a migration process framework composed of two processes for the redevelopment of the whole system from scratch and one process for creating new features with a microservice architecture on top of the existing system. 
They identified three different processes for migrating from a monolithic system to a microservices-based one. The goal of the first two processes is to support companies that need to migrate an existing monolithic system to microservices by re-implementing the system from scratch. The aim of the third approach is to implement new features only as microservices, to replace external services provided by third parties, or to develop features that need important changes and therefore can be considered as new features, thus gradually eliminating the existing system. 
All three of the identified processes are based on four common steps but differ in the details. 

\begin{itemize}
    \item \textit{Analysis of the system structure}. All processes start by analyzing dependencies mainly with the support of tools (Structure101, SchemaSpy~\footnote{http://schemaspy.sourceforge.net/}, or others) 

\item	\textit{Definition of the new system architecture}. Architectural guidelines or principles, and proposal of a decomposition solution into small microservices are defined. The decomposition is always done manually. 
\item	Prioritization of feature/service development. In this step, all three processes identify and prioritize the next microservices to be implemented. Some processes prioritize microservices based on customer value; others according to components with more bugs; and yet others prioritize the development of new features as microservices, expecting that, in the long run, the new ecosystem of microservices will gradually replace each feature of the existing monolith.  
\item	\textit{Coding and Testing} are then carried out like any other software development project. Developers adopt the testing strategy they prefer. However, in some cases, testing of the different microservices is performed by doing unit testing at the microservices level and black-box testing at the integration level. 
\end{itemize}

In this work, we focus mainly on the first two steps, supporting companies in the analysis of the system structure and in the identification of decomposition alternatives. The architectural guidelines should be defined by the company based on their internal policies. 

\subsection{Architectural Goals}
The microservices should be as cohesive and decoupled as possible ~\cite{fowler2014microservices}.
The motivation of such architectural characteristics is to keep the maintenance as local as possible. In other words, the changes to the source code should be local to one microservice. Such \textit{decoupled} architecture also supports independent development and deployment of the microservices.
Sam Newman~\cite{Newman2015} describe loose coupling as follows: “a loosely coupled service knows as little as it needs to about the services with which it collaborates”. 

% \kari{TODO: check overlapping with text in next section} @Davide. From my point of view, it is ok to introduce the theory here, and instantiate then in practice in Sect. 3
\textit{Cohesion} is related to decoupling and measures the degree to which the elements of a certain class belong together. Cohesion measures how weakly the functionalities of different modules are related to each other~\cite{Fenton2014}. High cohesion often relates to low coupling ~\cite{Kramer2004}~\cite{Jabangwe2015}. If the components of the software have high cohesion, the reasoning of the system is easier ~\cite{Kramer2004}. Thus, high cohesion supports efficient development and maintenance of the system.

In the design of microservice-based systems the developers target at high cohesion and low coupling by grouping the functionality and components according to the business processes.  Then, changes to a functionality should lead changes to one microservice only~\cite{Newman2015}.

Because cohesion and decoupling are key qualities of microservices, the dependency information is needed in the decomposition process. The commonly used dependency analysis tools, such as Structure 101, are based on static analysis of dependencies. They do not know which inter-component calls are really made and they do not recognize the full call paths. Our approach uses the dynamic dependency information that a process mining can provide.
The mining provides recommendations, and analysis can then be used for reasoning. At this point, we do not aim at completely automated decomposition.
In the next Subsection we report the underlying assumptions of our approach and details of the decomposition process.

\subsection{Decomposition Framework Assumptions}
The core assumption of our approach is existence of an extended log trace that has been collected at runtime. This means that the whole chain of operations after any external trigger can be traced from the log files.
Examples of such external events include any user operations (e.g., clicking on a button) and calls from other applications (e.g., APIs or command line).
The log file must include information about all methods and classes involved in serving of the request. 
The complete execution path must be completely traceable from the entry point to the access to the database (if any) and to the results returned to the client. The log must also include the start and end events. A hypothetical example of the data reported in the log file is shown in Table \ref{tab:Trace}.  A trace in Table~\ref{tab:Trace} is identified by a session ID. That ID distinguishes the trace from other sessions.

\begin{table}[]
\small\centering
\begin{tabular}{p{1cm}|p{1cm}|p{1cm}|p{1.4cm}|p{1.4cm}}
\hline
\textbf{Start Time}	& \textbf{End Time}	& \textbf{Sess.ID} &
% \textbf{Activity}	&
\textbf{Class}	& \textbf{Method} \\\hline
00:00	&00:36	&S1	&Form.jsp& 	btnClick()\\
01:00	&01:39	&S1		
&A.java	& a()\\
01:40	&01:45	&S1	&A.java	&b()\\
01:45	&01:55	&S1	&	B.java	&b()\\
01:56	&02:05	&S1	&B.java	&c()\\
02:05	&02:13	&S1	&DB.java&	query()\\
02:14	&02:21	&S1	&DB	&TABLE A\\
02:22	&03:28	&S1	&DB	&TABLE B\\
02:29	&02:36	&S1	&B.java	&c()\\
02:36	&02:45	&S1	&B.java	&b()\\
02:46	&02:55	&S1	&A.java	&b()\\
02:56	&03:03	&S1	&A.java	&c()\\
03:04	&03:16	&S1	&Results.jsp	&render()\\
 \hline
\end{tabular}
\vspace{5mm}
\caption{Example of Log Traces (Timestamps are shortened for reasons of space)}
\label{tab:Trace}
\end{table}

There are several ways to collect the traces. 
One possible method is to instrument the source code, but using Aspect-Oriented Programming (AOP) can be done too, like in the work done by Suonsyrjä~\cite{Sampo}.
For some runtime systems it is also possible to instrument the executable file with tools like  Elastic APM~\footnote{The Elastic APM Libraries. https://www.elastic.co/solutions/apm}.
For Java programs our current recommendation is to use Elastic APM since the instrumentation with it requires a minimal effort. 
Depending on the language and on the technology adopted, other tools such as  Dynatrace\footnote{Dynatrace https://www.dynatrace.com} or Datadog\footnote{Datadog https://www.datadoghq.com} could be also used.  
% \kari{Speculate about specific tool that also supports step 1}

\section{The Microservice Measurement Framework}
\label{sec:measurementFramework}

In this Section, we propose our microservice measurement framework. The framework has the goal of supporting companies to compare different microservices-based solutions, but also to understand the high-level architectural quality of their current system. 

The measurement framework is based on availability of log files from real execution and is composed of four measures: coupling (CBM), number of classes per microservice (CLA), number of duplicated classes (DUP), and frequency of external calls (FEC).
% The ideas of the first three measures were already presented in \cite{DavideKari2019} the fourth is a new measure. \kari{Should we keep this sentence.}

This measurement framework is used on the decomposition framework presented in Section \ref{sec:Framework}.

\subsection{Coupling (CBM)}
As reported in Section 2.3, in successful decompositions the coupling between microservices should be minimized and cohesion should be maximized. A comprehensive calculation of these measures would require knowledge of information beyond the log traces -- for instance knowledge about access to local variables. Thus, we need to rely on approximation.  
%The decomposition to microservices should minimize coupling and maximize cohesion. Coupling and cohesion can be calculated with different approaches. While coupling can be obtained from our log traces, for all the cohesion measures we also need to know about the access to the local variables of each class, which makes it impossible to calculate them from the data reported in the log traces.
One way to approximate coupling is to estimate it as an inverse to cohesion. Coupling can be considered as inversely proportional to cohesion and therefore, a system with low coupling will have a high likelihood of having high cohesion~\cite{Jabangwe2015}. 
%However, coupling is commonly considered as inversely proportional to cohesion~\cite{Jabangwe2015}. Therefore, a system with low coupling will have a high likelihood of having high cohesion ~\cite{Jabangwe2015}. 

In our framework we adopt the metric ''Coupling Between Microservice'' (CBM)~\cite{DavideKari2019},  a coupling measure inspired by the well-known Coupling Between Object (CBO) metric proposed by Chidamber and Kemerer~\cite{Chidamber1994}.
%We define the Coupling Between Microservice (CBM) extending the well-known Coupling Between Object (CBO) metric proposed by Chidamber and Kemerer~\cite{Chidamber1994}.
CBO counts the number of classes coupled with a given class. Classes can be coupled through several mechanisms, including method calls, field accesses, inheritance, arguments, return types, and exceptions.
% \kari{I wonder how close to CBO we actually are. Our measure is quite different.} Not really close, but it is inspired by CBO... 
%CBO represents the number of classes coupled with a given class (efferent couplings and afferent couplings). This coupling can occur through method calls, field accesses, inheritance, arguments, return types, and exceptions.
\newpage
In \cite{DavideKari2019} we calculate the relative CBM for each microservice as follows: 
%We calculate the relative CBM for each microservice as follows: 
\begin{center}
\[\text{CBM}_{MS_{j}} = \frac{\text{Number of external Links}}{\text{Number of Classes in the Microservice}}\]
\end{center}

In this formula “Number Of External Links” represents the number of call paths to external services. So, external services that are called several times, even by different classes of the microservice, are only counted once. The external services could be other microservices or services external to the whole system. The call frequency of external calls should also be take into account, but we have separate measure presented in Subsection \ref{sub:fec} for that. 
%where “Number Of External Links” represents the number of calls to external services used in each class of microservice. An external service linked several times by different classes of the same microservice is only counted once. External services could be other microservices, external APIs, etc. 

% \noindent
% CBM is calculated independently for each microservice and presented in a table for step 6. 

\subsection{Number of classes per microservice (CLA)}

This measure is an abstraction of the size of the microservice, and allows the developer to discover services that are either two big or too small compared with other microservices. In general, smaller microservices are easier to maintain and thus large microservices should be avoided.

In some cases optimizing for CLA measure leads to compromises to other measurements. For instance, larger number smaller microservices may lead to stronger coupling (CBM) and higher frequency of external calls (FEC). 

%Thus, this measure is also used in Step 6.
%This measure helps to understand how big the microservice identified is and to identify if there are microservices that are too big compared to others. the number of classes should be minimized since the smaller the number of classes the more independent its development can be. Considering the example reported in Figure~\ref{fig:Fig3}, the decomposition option 1 has 7 classes while option 2 has six classes.

\subsection{Number of duplicated classes (DUP)}
The execution traces often have common sub-paths, i.e., some
classes and methods are common for several execution traces.
If traces should be implemented in different microservices, one way to increase independence is to is duplicate part of the code to several microservices. For example, method j of class E (Figure~\ref{fig:Fig3}) is used by two execution traces. In that example the decomposition option 1 has one duplicated class, while option 2 requires no classes to be duplicated.
%In some cases, several classes will be in common between two execution traces. As example, the method j in Class E (Figure~\ref{fig:Fig3}) is used by two execution traces. In the example depicted in Figure~\ref{fig:Fig3}, decomposition option 1 has one class that needs to be duplicated, while option 2 requires no classes to be duplicated. 
%
%This measure is also used to evaluate the different slicing options in step 6.
Duplicated classes increases size of the system and complicates the maintenance process.
%This measure helps to reason about the different slicing options, considering not only the size of the microservices but also the number of duplications, that will be then reflected in the development of the microservices.  
%Duplicated classes should be avoided since the duplication adds to the size of the system and its maintenance.

\subsection{Frequency of external calls (FEC)}
\label{sub:fec}
Calls between microservices are computationally substantially heavier than calls within a microservice. Thus, reducing of the frequency of external calls optimizes the performance and delays. Since our approach is based on log-file analysis, we have the frequency information available. 

We use the call frequency presented in \ref{tab:Freq} as an input to a measure relative Frequency of External Calls (FEC):

\begin{center}
\[\text{FEC}_{MS_{j}} = \frac{\text{Number of Call Instances}}{\text{Number of Classes in the Microservice}}\]
\end{center}

As an example consider execution paths and decomposition options in Figure \ref{fig:Fig3}. For the sake of the example we assume that
\begin{itemize}
    \item Path A.a() $\rightarrow$ A.b() $\rightarrow$ B.c()  $\rightarrow$ B.d() is called 200 times
\item Path C.e() $\rightarrow$ C.f() $\rightarrow$ D.g()  $\rightarrow$ D.h() is called 200 times.
\item Path C.e() $\rightarrow$ C.f() $\rightarrow$ F.j() $\rightarrow$ D.g()  $\rightarrow$ D.h() is called 50 times
\item
Path E.i() $\rightarrow$ E.j() $\rightarrow$ F.k()  $\rightarrow$ F.l() is called 100 times.
\end{itemize}
With the input data we can calculate the number of internal calls, external calls, and FEC per each microservice. In table \ref{tbl:fecmeasures} we also show the total number of internal in  microservices (internal C), total number of calls between microservices (external C) and relative computational load (load). In this example we assume that an external call is 1000 times heavier than an internal call.
\begin{table}[]
\small\centering
\begin{tabular}{|l||c|c|c||c|c|c|}
\hline
MS split & internal c & external c & load & $FEC_{MS1}$ & $FEC_{MS2}$ & $FEC_{MS3}$ \\
\hline
0: A+B, C+D, E+F & 1150 & 100 & 101550 & 0 & 25 & 25 \\
1: A+B, C+D+E.j, E+F & 1650 & 0 & 1650 & 0 & 0 & 0 \\
2: A+B, C+D+E+F & 1650 & 0&  1650 & 0 & 0 & \\
\hline
\end{tabular}
\caption{Example analysis of call frequencies between microservices}
\label{tbl:fecmeasures}
\end{table}

\section{The Decomposition Framework}
\label{sec:Framework}

%Applications built from microservices should be as decoupled and as cohesive as possible ~\cite{fowler2014microservices}.
%In the case of loosely coupled services, changes to one service should not require changes to other services. Therefore, the developers of microservices can change and deploy their microservices independently. 
%As reported by Sam Newman~\cite{Newman2015}, “a loosely coupled service knows as little as it needs to about the services with which it collaborates.”. 
%Therefore, developers should limit the number of different types of calls from one service to another. 

\begin{figure*} [!ht]
    \centering
    \includegraphics[width=1.0\columnwidth]{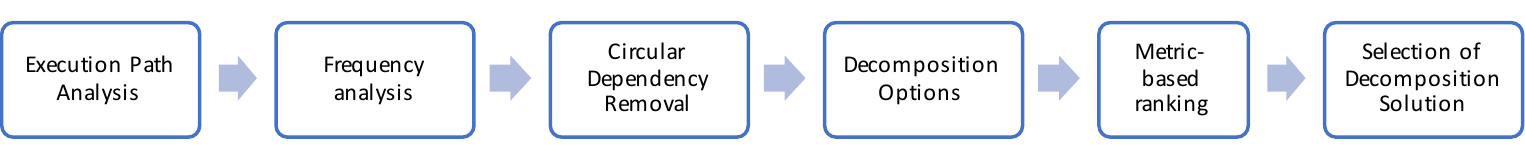}
    \caption{The Decomposition Process (From~\cite{DavideKari2019})}
    \label{fig:Dcomposition}
\end{figure*}

In this Section, we describe a decomposition framework that uses the data from the execution path analysis to discover the optimal split to micro services. A top-level description of the framework is given Figure \ref{fig:Dcomposition}. 

When the log files are available, the decomposition process defined in (Figure~\ref{fig:Dcomposition}) can be started. The process consists of six steps that are outlined in the following subsections.
%Once the log files are created, companies can start the decomposition following our 6-step process (Figure~\ref{fig:Dcomposition}).

\subsection{Step 1: Execution Path Analysis.}
\begin{figure} []
    \centering
    \includegraphics[width=0.5\columnwidth]{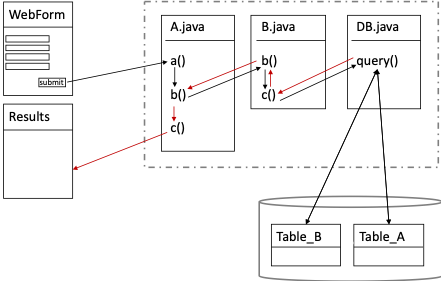}
    \caption{Simplified Process Example (From~\cite{DavideKari2019})}
    \label{fig:Fig2}
\end{figure}

As our approach aims to optimize the system for the often used call sequences, the first step of our approach is to identify the most frequently executed call paths from the log files.
One way to do that is use of a process-mining tool.
In our case, we used DISCO~\footnote{https://fluxicon.com/disco/} to graphically represent the business processes found in the log files.
%In the first step, we identify the most frequently used execution paths with a process-mining tool. In our case, DISCO~\footnote{https://fluxicon.com/disco/} was used to graphically represent the business processes by mining the log files.
Other similar tools can be used instead.
%The same result can be obtained by any other alternative process-mining tool.
The outcome of step 1 is a graphical representation of the processes and call graphs. One example of such graphical diagram is presented in Figure~\ref{fig:Fig2}. The diagram shows the call paths between classes, methods and data bases with arrows.
This figure provides the user with the following information:
%The result is a graphical representation of the processes, reporting each class and database table used in the business processes, with a set of arrows connecting each class based on the log traces. 
%The result of this first step produces a figure similar to the one presented in Figure~\ref{fig:Fig2}, that allows to understand:
\begin{itemize}
    \item The actually executed call paths in the system. Possible but never executed paths are not shown in this figure. 
%    \item Runtime execution paths of the system. Paths never used, even if possible, are not represented in the figure. 
    \item Inter-class dependencies in the system. The dependencies are visualized with arrows between methods and classes. The external dependencies to libraries or web-services are also visualized.
    %\item Dependencies between the classes of the system. The arrows represent the dependencies between methods and classes. External dependencies to libraries or web-services are also represented. 
    \item The usage frequency of each path. Process mining tools may present the frequency with thickness of the arrows or in a separate table as in Table~\ref{tab:Freq}. 
    %\item The usage frequency of all paths. Process mining tools present the most used processes with thicker arrows.
    \item Branches and circular dependencies. If the system has circular dependencies or branches in the call path, those can easily be found from the visualization.
    %\item Branches and Circular Dependencies. The graphical representation allows easy discovery of circular dependencies or branches (e.g., conditional statement that led to different path based on the input provided), in case they exist.
\end{itemize}

The call paths, shown with chains of arrows in Figure~\ref{fig:Fig2}, form candidates for business processes that are later used in the decomposition to microservices. 
For example, the path documented in Table~\ref{tab:Trace} is visualized in a business process shown in Figure~\ref{fig:Fig2}.

%The complete chain of arrows forms a candidate of a process. Figure ~\ref{fig:Fig2}.  represents a simplified example of one business process representing the data reported in Table~\ref{tab:Freq}. 

\begin{table}[]
\small
\centering
\begin{tabular}{p{7.5cm}|p{1cm}}
\hline
\textbf{Path} & \textbf{Freq.} \\\hline 
A.a(); A.b(), B.b(), C.c(), DB.query, Table A, Table B, … & 1000 \\
A.b(); A.c(), B.a(), C.c(), DB.query, Table A, Table B, … & 150 \\ \hline
\end{tabular}
\caption{Frequency analysis of each execution path (From~\cite{DavideKari2019})}
\label{tab:Freq}
\end{table}

\begin{figure} []
    \centering
    \includegraphics[width=\columnwidth]{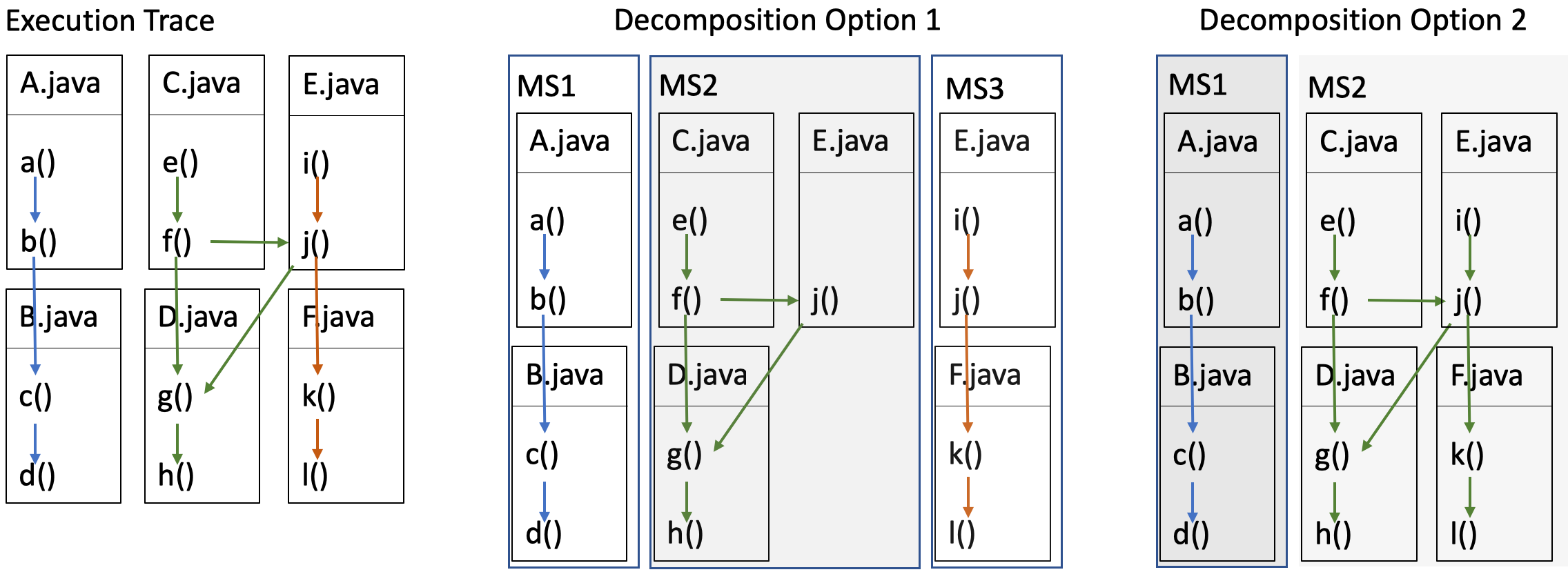}
    \caption{Simplified Process Example (From~\cite{DavideKari2019})}
    \label{fig:Fig3}
\end{figure}

\subsection{Step 2: Frequency Analysis of the Execution Paths.}
In our approach, the call frequency is a major contributor for the produced recommendations. Thus the frequency should be studied and analyzed.
For visual inspection, the process analysis tools can provide help. For instance, in the DISCO tool, the thickness of the arrows reflect the call frequency. In addition to visual inspection, we use concrete numeric data for all execution paths with their usage frequency.
So, the output of this step is a table similar to Table~\ref{tab:Freq}.
%The thickness of the arrows created by the DISCO tool indicates the frequency of the calls between classes. This makes it possible to clearly understand which execution path are used most frequently and which classes are rarely or never used during runtime. The output of this step is a table representing all the different execution paths with the frequency of their usage. 

The Frequency Analysis step helps the architect to select the potential decomposition options. The numbers are used to calculate the measures presented in Section \ref{sec:measurementFramework} and used in step 6 (see Subsection \ref{sub:metric-ranking}).

\subsection{Step 3: Removal of Circular Dependencies}
% \kari{TODO: need to be discussed with Davice}
In this step, we first find circular dependencies by analyzing the execution paths reported in the table generated in the first Step (e.g. Table~\ref{tab:Freq}). This can be done with a simple algorithm to discover cycles in the execution paths. 
In the case of circular dependencies, software architects should discuss with the development team how to break these cycles. One example of the patterns that can be applied to break the cycles is Inversion of Control~\cite{Martin2003}. However, every cyclic dependency could need a different breaking solution that must be analyzed carefully.
The result is a refined version of the execution path table (see Table~\ref{tab:Freq} as example).  

Although  this step is not part of the decomposition, it is important to consider cyclic dependencies to avoid possible  deadlocks
% \kari{Is it? Depends on how calls between microservices are implemented?} between microservices
, and to better architect the system. 

\subsection{Step 4: Identification of Decomposition Options}
In this step the execution paths from Step 3 are used to identity different decomposition opportunities.
Visual inspection of the call graphs like those shown in Figure~\ref{fig:Fig3} is used.
%Starting with the execution paths without cyclic dependencies obtained from Step 3, we identify different decomposition alternatives by visually inspecting the generated graphs. 
%The candidate processes may have common sub-paths, i.e., the processes may merge or split. Thus, different decomposition solutions are possible.
We have relied on manual expert-based decomposition, but an algorithm to create different decompositions could be developed.
%This process could also be automated by developing an algorithm that provides all different decompositions based on the paths with fewer intersections. However, in this case, we rely on the expert-based decomposition. 

% \kari{changes in the following go beyond wording changes}
The execution paths may merge to common sub-paths or split into several branches. This leads to alternative decomposition solutions. This is demonstrated in Figure~\ref{fig:Fig3}.
If the six source files each providing implementation of a class
% \kari{I wonder if the figure should use "class A" instead of "A.java"?} 
are assigned to three different microservices as A.java+B.java, C.java+D.java and E.java+F.java, the calls form C.f() to E.j() and E.j() to D.g() are inter-service calls. These calls are heavier operations than local calls and expand the external interfaces of the microservices. If use of external calls is not feasible, there are two other alternatives that can be proposed. The first option is to use three microservices so that class E (or at least function j() of it) is duplicated in two microservices. The other option is to decompose into two microservices as shown in the rightmost part of Figure~\ref{fig:Fig3}. Obviously, there is also the alternative to allow external calls and have three microservices with no duplications.
%As highlighted in Figure~\ref{fig:Fig3}, the decomposition options need to deal with the duplication of some classes or methods. As example, the execution traces reported in Figure~\ref{fig:Fig3} show that both the green and the orange execution traces use j().
%Therefore, software architects could propose two decomposition alternatives. The first option includes the creation of three microservices where class E.java() is duplicated in microservice MS2 and MS3. The second option includes the creation of two microservices, merging MS2 and MS3. 

All these options have their advances and disadvantages, and the team should discuss the alternatives is from multiple view points. The consideration could include both functionality of the software -- if the paths belong logically together, and development process -- what are the consequences of duplication to the development team(s).
In addition, the call frequency has to be taken into account. For example in above split and merge case, the team has to consider both the development effort and run-time cost of making the two call external. The metrics discussed \ref{sec:measurementFramework} help in analysing the run-time costs.

%Both options have pros and cons, but the decision of merging two execution traces or splitting into different microservices must be discussed with the team. If two microservices candidates for the splitting have different purposes, it is reasonable to consider the splitting. If they are doing the same thing, then it would be better to merge them into one single microservice.

\subsection{Step 5: Metric-based ranking of the decomposition options }
\label{sub:metric-ranking}

In this step, we apply the measures identified in the Measurement Framework (Section~\ref{sec:measurementFramework}), to help software architects to assess the quality of the decomposition options. 
%In this step, we identify three measures to help software architects to assess the quality of the decomposition options identified in Step 4: Coupling, Number of classes per microservices, Number of classes that need to be duplicated. 

Sometimes optimization of the measures contradict with each other. Currently, we propose use of judgment of the team, but in the future approaches like Pareto\cite{deb99} optimization could be used.

\subsection{Step 6: Selection of the decomposition solution }
\label{subsec:selection}
In final step, the decomposition alternatives identified in Step 4 and the measures collected in Step 5 are used by the software architects to decide which solution to take. %An example summary of the data is shown in Table~\ref{tbl:measures}. 
% \kari{Or do we just rely on table 6 and simplify this?}. 

%The following has been covered already
%In this example, the all three decomposition solutions have five microservices and for each service we have to calculate the four measures:
%\begin{itemize}[leftmargin=1.5cm]
%    \item[CBM] Coupling between microservices
%    \item[CLA] Total number of classes
%    \item[DUP] Number of duplicated classes in a microservice
%    \item[FEC] Relative frequency of external calls
%\end{itemize}

%This is the final step where, based on the different decomposition alternatives identified in Step 4, and on the measures collected in Step 5, software architects can decide which solution to adopt by merging or splitting existing processes. 
Our approach does not automatically rank the solutions to any order. The software architects should consider the provided recommendations and measurements before selecting the solution. The team should discuss the relevance of the differences in their case.

\section{Discussion}
\label{sec:discussion}
In this work we proposed a microservice measurement framework and we applied in in our previously proposed decomposition process~\cite{DavideKari2019}.
% to slice monolithic systems into microservices based on their runtime behavior. The earlier version of the decomposition has been evaluated in an industrial context \cite{DavideKari2019}. We have also created measurement framework that assist developers in evaluation of microservice architectures.

The measurement framework is based both on static and dynamic measures collected at runtime. 
The main benefit of analyzing runtime information is the availability of the data on the usage of each component, together with the dynamic analysis of dependencies. %We identified several dead methods and classes that were never used at runtime and we also spotted some cyclic dependencies. 
% The static analysis of dependencies would have spotted the circular dependencies but not all the dead code. 
The dynamic measures allow to better understand the behavior of the system, and to analyze the dynamic coupling. Moreover, thanks to the dynamic measures collected, such as the frequency of usage of each method, the software architects can better understood which features is used more, we prioritized the development and the slicing of the monolithic system differently. 
% Without the information on the frequency of usage of methods, we could have created a microservice that would have done most of the computational tasks. 

% We are aware about possible threats to validity. We tried to reduce them by applying a common process mining tool (DISCO), that has been developed for several years and has been adopted by several companies and universities. However, the tool could have identified some processes incorrectly. Moreover, we are aware about the complexity related to the data collection, since to adopt our process, companies need to instrument their code to collect the log files at the method level. About the generalizability of the results, the validation case study was based on an analysis of the processes of one company. The project manager and the software architect had a limited experience decomposing systems into microservices but the authors of this paper have more than four years of experience in supporting companies in decomposing systems into microservices and closely followed them during the migration.  

Companies could benefit from our lessons learned, by applying our proposed frameworks to decompose their monolithic system, but also to evaluation and monitoring the runtime behaviors or existing microservices to continuously understand possible issues. 
Moreover, the microservice measurement framework will allow software architects to clearly evaluate different decomposition options, with the usage of repeatable and objective measures. 

Despite this approach being very beneficial in our case company, the results could have a different impact on other companies. Researchers can benefit from this approach and extend it further. New optimization metrics could be defined, and in theory, it would be possible to propose an automated decomposition approach that would identify the slices by maximizing the metrics identified. Genetic algorithms could be a possible solution for this idea.

\section{Related Work}
\label{sec:related}
% \kari{Moved "related" to here (re-structuring was asked for the Springer edition)}

% \kari{Text starts with new content (three papers and continues then with old.}

Fritzsch et al~\cite{Fritzsch2018} present a classification for the refactoring approaches. Our approach should be categorized as \textit{Workload-Data aided} in this classification since we use operational data, i.e., dynamic data, in decomposition and analysis.

Bogner et al have conducted a literature review of maintenance metrics of microservices~\cite{Bogner2017}. The report summarizes several metrics that specialize in service-based systems instead of metrics designed for object-oriented systems. Although that research assume use of static info only, these metrics should be interesting for us in future research.

One case of refactoring a legacy system to a service based system has been reported buy Khadka et al. \cite{Khadka2013}. Their case has substantial similarities to our approach and case. They also stress the importance and difficulty of finding the right set of services. They also analyze the call paths to find hotspots in the code. However, they do not present a systematic and repeatable process for decomposition. Actually, only a limited set of research works propose systematic approaches for developers in decomposing their systems into an optimal set of microservices.  

Chris Richardson ~\cite{RichardsonCube2017} put scalability into focus and propose a decomposition approach based on the ''scalability cube'' where applications can be scaled based on the X, Y or Z axis. The X-axis and Z-axis scaling consists of running multiple copies of an application behind a load balancer. The Y-axis axis scaling is the real microservice decomposition approach, that propose to split the application into multiple, different services. Each service is responsible for one or more closely related functions. The decomposition is then based on two approaches: decomposing based on verbs used into the description of the service or decomposing by noun creating services responsible for all operations related to a particular entity such as customer management. Richardson also recommend to use combination of verb-based and noun-based decomposition when needed. 
 
% \kari{Would be nice write a sentence or two towards concreteness.}
% %Abbott and Fischer~\cite{Abbott2015} proposed a decomposition approach based on the ''scalability cube'', which splits an application into smaller components to achieve higher scalability.
Richardson~\cite{Richardson2017} also mention this approach in his two decomposition strategies:

%Richardson~\cite{Richardson2017} also mentioned this approach in his four decomposition strategies: 
\begin{itemize}
    \item ''Decompose by business capability and define services corresponding to business capabilities'';
    \item ''Decompose by domain-driven design sub-domain'';
\end{itemize}

In an older version of this page~\cite{Richardson2017} (2017), Richardson proposed other two patterns: 
\begin{itemize} \item ''Decompose by verb or use ‘cases’ and define services that are responsible for particular actions'';
    \item ''Decompose by nouns or resources by defining a service that is responsible for all operations on entities/resources of a given type''.
    \end{itemize}

% \kari{I did not touch this paragraph due to the above question} The first two strategies are mostly abstract patterns of human decisions~\cite{Yourdon1979} while the others are based on predefined criteria. 
Kecskemeti et al.~\cite{Kecskemeti2016} proposed a decomposition approach based on container optimization. The goal is to increase the elasticity of large-scale applications and the possibility to obtain more flexible compositions with other services.

% Arndt and Guercio \kari{What the heck is this? I guess it is not \url{https://ieeexplore.ieee.org/document/227904} } suggest decomposing a monolith system using a layered architecture style, with the outcome being highly cohesive and loosely coupled services, such as representation and business services. 
Another decomposition possibility is to start from a monolithic system and progressively move towards a microservices-based architecture~\cite{Zimmermann2017} or delivering separate microservices by splitting a development team into smaller ones responsible for a limited group of microservices. 
% \kari{Why are these two in same paragraph? The former is about architecture-based tactics while the latter is about evolutionary process.}

Vresk et al.~\cite{VreskC16} defined an IoT concept and platform based on the orchestration of different IoT system components, such as devices, data sources, data processors, and storage. They recommend an approach similar to the one proposed by Richardson's Scalability Cube~\cite{RichardsonCube2017} combining verb-based and noun-based decomposition approaches. The proposed approach hides the complexity stemming from the variation of end-device properties thanks to the application of a uniform approach for modeling both physical and logical IoT devices and services. Moreover, it can foster interoperability and extensibility using diverse communication protocols into proxy microservice components. 

Gysel et al.~\cite{Gysel16} proposed a clustering algorithm approach based on 16 coupling criteria derived from literature analysis and industry experience. This approach is an extensible tool framework for service decomposition as a combination of a criteria-driven methods. It integrates graph clustering algorithms and features priority scoring and nine types of analysis and design specifications. Moreover, this approach introduces the concept of coupling criteria cards using 16 different instances grouped into four categories: Cohesiveness, Compatibility, Constraints, and Communications. The approach was evaluated by integrating two existing graph clustering algorithms, combining actions research and case study investigations, and load tests. The  results showed potential benefits to the practitioners, also confirmed by  user feedback.

Chen et al.~\cite{Chen2017} proposed a data-driven microservices-oriented decomposition approach based on data flow diagrams from business logic. Their approach could deliver more rational, objective, and easy-to-understand results thanks to objective operations and data extracted from real-world business logic. Similarly, we adopt process mining to analyze the business processes of a monolithic system.

Alwis et al.~\cite{Alwis2018} proposed a heuristic to slice a monolithic system into microservices based on object subtypes (i.e., the lowest granularity of software based on structural properties) and functional splitting based on common execution fragments across software (i.e., the lowest granularity of software based on behavioral properties). This approach is the closer to our work. However, they analyzed the system by means of static analysis without capturing the dynamic behavior of the system and they did not propose measures to evaluate the quality of the slicing solution proposed.

Taibi et al.~\cite{TaibiIEEEsw}~\cite{TaibiBOOK}~\cite{TaibiBOOK1}, proposed a set of patterns and anti-patterns that should be carefully considered during the microservice decomposition~\cite{TaibiIEEEsw}~\cite{TaibiBOOK} recommending to avoid a set of harmful practices such as cyclic dependencies and hard-coded endpoints but also to consider critical anti-patterns and code smells~\cite{TaibiJL17} that can be generated into the monolithic system.

\section{Conclusions}
\label{sec:conclusions}
The decomposition of monolithic systems into microservices is a very complex and error-prone task, commonly performed manually by the software architect. 

In this work, we first proposed a new microservice measurement framework based on 4 measures: coupling, number of classes per microservices, number of duplicated classes and frequency of external calls. 
The goal of our framework is to support software architects to compare different microservice decompositions, by means of a set of objective and repeatable measures. 

We instantiated our measurement framework in the context of the previously proposed  process-mining decomposition approach~\cite{DavideKari2019}.

%  We demonstrated the usefulness of existing process-mining approaches for decomposing monolithic systems based on business processes identified from the process-mining approach and validated the new of measurement framework.

Our goal is not the automated slicing of monolithic systems but to provide extra support to software architect, to help them in identifying different slicing options reducing the subjectivity and to measure and compare the different solutions objectively. 

The microservice measurement framework can be adopted independently from the decomposition process used. As example, software architect might manually identify two decomposition options for a monolithic system. The measurement framework will support them in the  comparison of their decomposition options. 
% We validated our approach with an industrial case study. 

% As a result, the company reported that the decomposition process simplified the identification of alternative decomposition solutions, and the measurement framework helped to objectively evaluate the quality of the decomposition. Moreover, our process-mining approach keeps track of the classes and methods traversed by each process, which does not only help to identify business processes but also makes it possible to discover possible issues in the processes, such as unexpected behavior or unexpected circular dependencies. 

We recommend companies to apply periodically our measurement framework also in case of existing microservices-based systems. The historical analysis of the evolution of the system might provide useful information on the quality of the system and also be a trigger for future refactorings. 

% decomposition process if they already collected log data is available. Moreover, we also recommend companies adopting different decomposition approaches to measure their decomposition options with our framework to reduce the subjectivity of the manual evaluation. 

Future works include the validation of the framework, both in the context of manual decompositions and when using process-mining based approaches. Moreover, we want to evaluate the application of our approach in  development of a tool to facilitate the identification of the process, the automatic calculation of the metrics, and identification of other measures for evaluating the quality of the decomposition. We already started to develop a tool to automatically identify dependencies between microservices~\cite{microdepgraph} and we published a dataset containing the analysis of 20 projects~\cite{RahmanDataset2019}.

We are also planning to further empirically validate this approach with other companies and to include dynamic measures for evaluating the quality of the system at runtime~\cite{LenarduzziECISM2017}~\cite{TosiDynamic}. In the future, we are also planning to adopt mining software repositories techniques to identify the areas that changed simultaneously in the past, to help developers to understand pieces of code connected to each other. 

Another possible future work is to include identification of partial migration, i.e., migration of a limited set of processes from a monolithic system. Finally, we are also considering to extend this work by proposing not only different decomposition options but also a set of patterns for connecting microservices based on existing common microservices patterns ~\cite{Newman2015}~\cite{TaibiCLOSER} and anti-patterns~\cite{TaibiIEEEsw}\cite{TaibiBOOK}\cite{TaibiBOOK1}.

%\vfill
% \section*{\uppercase{Acknowledgements}}

% \noindent If any, should be placed before the references section
% without numbering. To do so please use the following command:
% \textit{$\backslash$section*\{ACKNOWLEDGEMENTS\}}

\bibliographystyle{splncs04}
\bibliography{example}

\end{document}